\documentclass[sigconf]{aamas}  

\usepackage{booktabs}

\setcopyright{ifaamas}  
\acmDOI{doi}  
\acmISBN{}  
\acmConference[AAMAS'18]{Proc.\@ of the 17th International Conference on Autonomous Agents and Multiagent Systems (AAMAS 2018), M.~Dastani, G.~Sukthankar, E.~Andre, S.~Koenig (eds.)}{July 2018}{Stockholm, Sweden}  
\acmYear{2018}  
\copyrightyear{2018}  
\acmPrice{}  

\begin{document}

\title{Valuing knowledge, information and agency in Multi-agent Reinforcement Learning: a case study in smart buildings}  




%
\author{Hussain Kazmi}
\affiliation{%
  \institution{KU Leuven and Enervalis}
  \city{Leuven} 
  \country{Belgium} 
}
\author{Johan Suykens}
\affiliation{%
  \institution{KU Leuven}
  \city{Leuven} 
  \country{Belgium} 
}
\author{Johan Driesen}
\affiliation{%
  \institution{KU Leuven}
  \city{Leuven} 
  \country{Belgium} 
}

\begin{abstract}  
Increasing energy efficiency in buildings can reduce costs and emissions substantially. Historically, this has been treated as a local, or single-agent, optimization problem. However, many buildings utilize the same types of thermal equipment e.g. electric heaters and hot water vessels. During operation, occupants in these buildings interact with the equipment differently thereby driving them to diverse regions in the state-space. Reinforcement learning agents can learn from these interactions, recorded as sensor data, to optimize the overall energy efficiency. However, if these agents operate individually at a household level, they can not exploit the replicated structure in the problem. In this paper, we demonstrate that this problem can indeed benefit from multi-agent collaboration by making use of targeted exploration of the state-space allowing for better generalization. We also investigate trade-offs between integrating human knowledge and additional sensors. Results show that savings of over 40\% are possible with collaborative multi-agent systems making use of either expert knowledge or additional sensors with no loss of occupant comfort. We find that such multi-agent systems comfortably outperform comparable single agent systems.
\end{abstract}

%


\maketitle


\section{Introduction}

The theoretical allure of reinforcement learning (RL) as an end to end black box method is obvious. By translating sensory input directly into meaningful control actions, robust optimal systems can be developed in a cost-effective way \cite{RL03}, \cite{RL01}, \cite{RL02}. In practice however, numerous trade-offs have to be made between quality of control and the cost associated with accomplishing it. These include (1) the extent of information available via sensors, (2) the level and ease of integration of prior human knowledge and (3) the possibility of deploying multiple agents to accelerate learning.

These trade-offs are embodied in smart buildings and smart grids, where agents are deployed to perform automated optimal control. The objective for control can vary from case to case but two common ones are to minimize overall energy consumption \cite{Energy} and peak power consumption \cite{Power}, while maintaining predefined user comfort bounds. Reducing energy consumption is a local objective where multiple agents act independently, since energy consumed in one household does not affect another. Peak shaving, on the other hand, is a multi-agent problem where different agents have to coordinate their energy consumption to reduce simultaneous demand.

In this paper we focus on optimizing energy consumption for hot water production, a load that is responsible for well over 10\% of the total energy consumed in modern residential buildings \cite{DHW}. We show that while each agent can act independently to optimize its load as explained in existing literature \cite{Energy}, \cite{Energy2}, a coordination mechanism to improve state-space exploration can substantially improve overall efficiency. This collaboration makes use of the insights developed in \cite{MARL}. 

Optimizing the hot water system in such settings can be considered an n-player finite, non-zero sum game of hidden information. Here, n-player refers to the fact that individual agents are operating in multiple houses in parallel to optimize their respective rewards. The overall problem is non-zero sum since an agent\textquotesingle s strategy does not directly affect other agents or their rewards. Hidden information refers to the fact that in most hot water systems sensing is limited to only a single temperature sensor which is not representative of the system.

The framework for optimizing hot water production presented next explores the trade-offs in RL mentioned at the beginning of the paper. Concretely, we investigate and compare the quality of end-to-end control learned using RL for hot water production in smart building communities which employ the same thermal equipment (e.g. in large apartment blocks and social houses etc.). To quantify the trade-offs highlighted earlier, we do this for different configurations using additional sensing, human domain knowledge and multiple agents.

\section{Methodology}
To integrate these three components, we first define a Markov Decision Process (MDP): \textit{M} = \{\textit{S}, \textit{A}, \textit{T}, \textit{R}\}. The structure of the MDP derives from the interactions between a hot water storage vessel, a heating element and the human occupant. The RL agent sends reheat commands to the vessel (via the heating element) following a policy, $\pi$, that minimizes energy consumption while maintaining occupant comfort. The control actions, $a_t$ $\epsilon$ $A$, are thus binary, and the reward stream $R(.)$ that the agent receives is a function of the energy consumed and the impact on occupant comfort. The vessel state, $S$, is given by a temperature distribution profile and is representative of the energy content in the vessel. The transition function, $T(.)$, defines the next state of the vessel as a function of current state, the agent\textquotesingle s action and stochasticity arising from human occupant behavior. The interactions of the agents with the storage vessel are simulated using a model fit to empirical data while occupant behavior is modelled as a stochastic time series fit using real world human behavior. The consumption time series are strongly auto- and cross-correlated. To investigate the aforementioned trade-offs, we consider variations involving the following:

\textbf{Information (I)}: In the default configuration, the storage vessel is equipped only with a mid-point temperature sensor. This is not enough to generalize because the temperature distribution is nonlinear and exhibits stratification effects \cite{Energy}. Additional sensors can facilitate learning of the distribution.

\textbf{Knowledge (K)}: There are two ways human knowledge can be encoded as prior knowledge for the agents: feature engineering based on available sensor data and constraining the behavior of the vessel model based on thermodynamic laws. This latter can include defining gradients and end point limits on the temperature distribution, its stratification and possible phase inversion. 

\textbf{Agency}: The structure of the MDP, M, is replicated across all households which share the same thermal equipment. Since different agents are driven to different regions of the state-space as a result of different human interactions, learning a shared representation of the transition function from sensor data can help in generalization by decoupling stochastic human behavior from deterministic storage vessel behavior. This also makes targeted exploration of the state-space a viable alternative to $\epsilon$-greedy strategies.

\section{Results and discussion}

\begin{figure}
\includegraphics[height=1.3in]{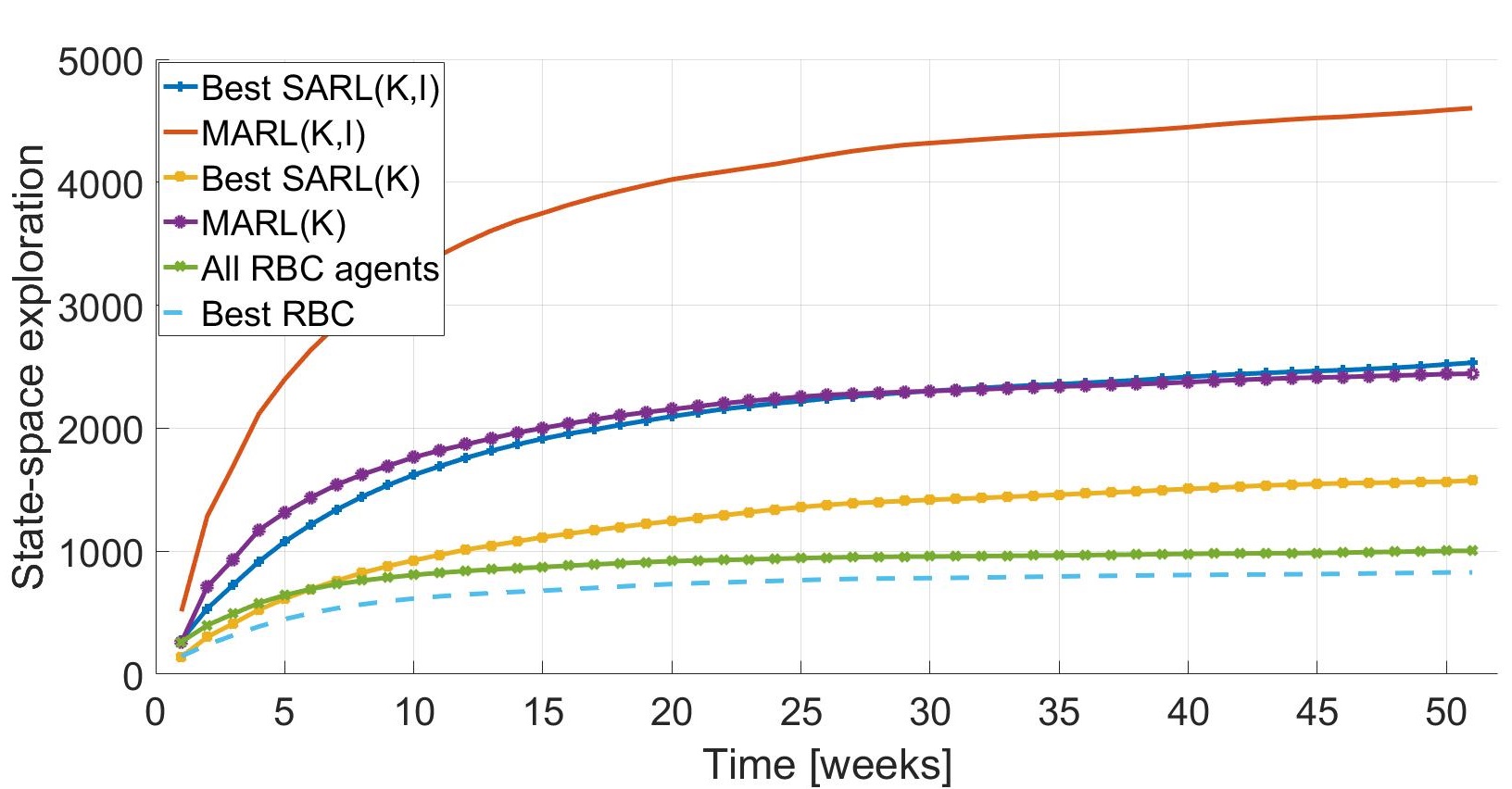}
\includegraphics[height=1.3in]{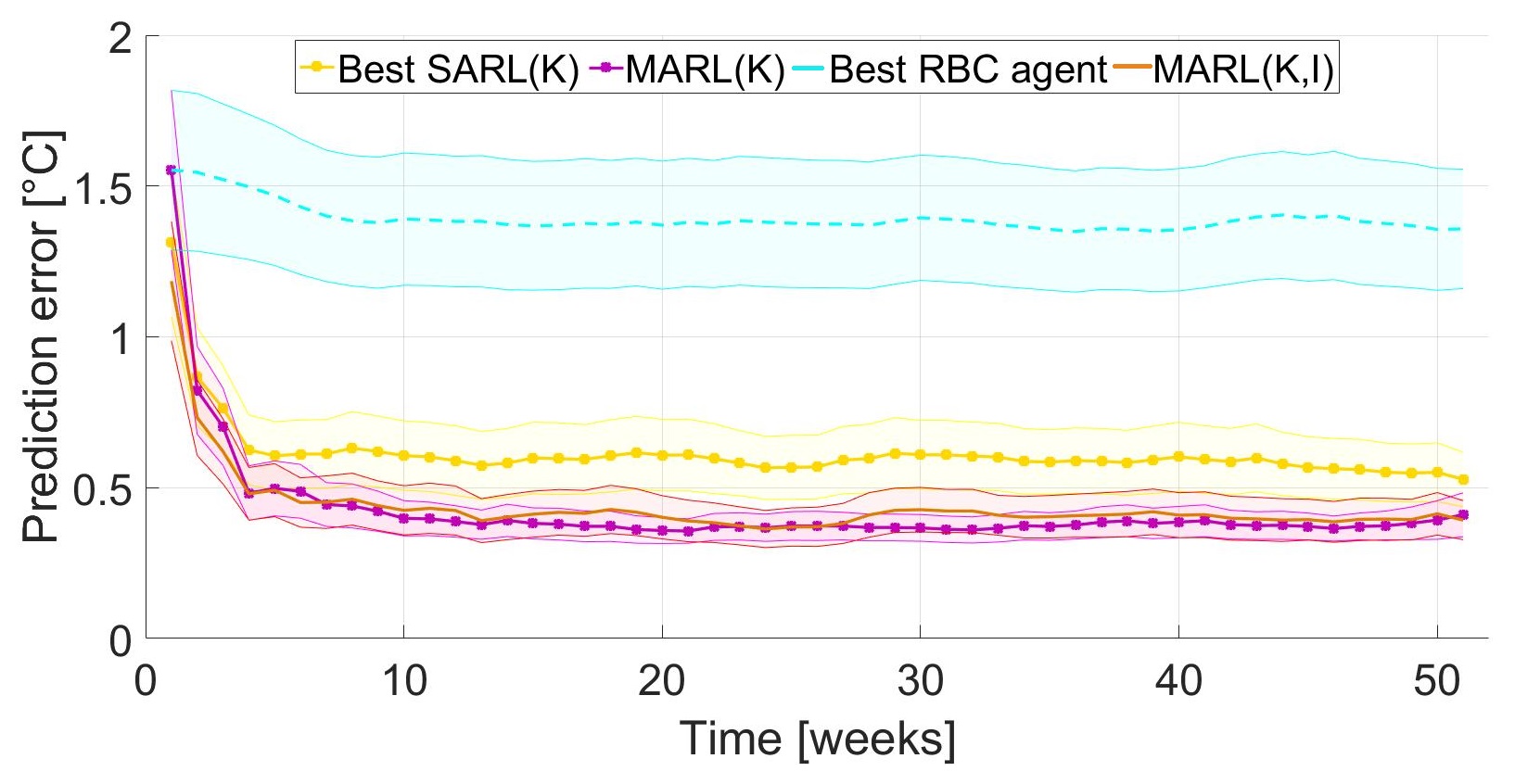}
\caption{(a) State-space exploration; (b) MAE for learnt transition model, with different configurations}
\end{figure}

Given a reasonably flexible algorithm, the amount of state-space exploration by the agent (if search is suitably diversified) is positively correlated with its generalization. This exploration is illustrated in Fig. 1a with different control strategies which include na\"ive rule based controllers (RBC) and their aggregation as well as the best single agent reinforcement learner for each household (SARL(K), where K represents domain knowledge), and aggregation of all such agents with targeted exploration (MARL(K)). Finally, we also consider both SARL(K) and MARL(K) augmented with extra sensing information (I): SARL(K,I) and MARL(K,I). 

It is evident from Fig. 1a and 1b that simply increasing the number of agents without adopting a more complex control policy and incorporating domain knowledge does not help the agent explore the state-space (see also Fig. 2). The exploration potential of such strategies also tapers off as the auto-correlated occupant behavior results in the agent visiting the same states repeatedly. This is reflected in Fig. 1b where the prediction MAE is uniformly high for strategies which explore less. An exception to this is MARL(K,I) which explores the most but has a performance no better than MARL(K); this is because additional sensing provides similar information as human expert knowledge. Fig. 2 reveals that both MARL(K) and MARL(K,I) learn reliable transition functions within two months, a feat that naive aggregation of RBC agents is unable even to after a year. The single agent, SARL(K) configuration suffers from prediction errors mostly at the transition between hot and cold water which has negative implications for the end user.

\begin{figure}
\includegraphics[height=2in]{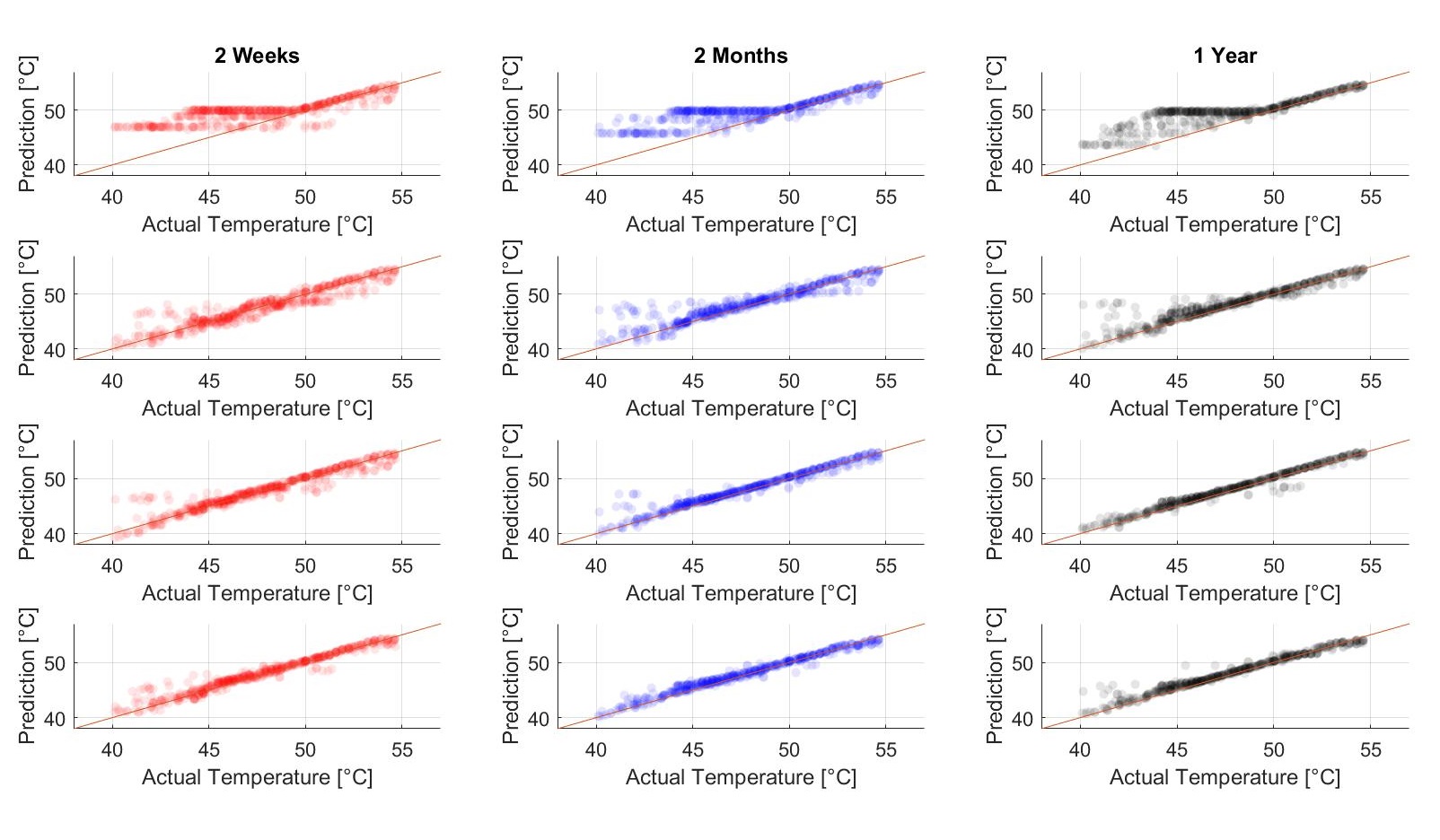}
\caption{Predicted and observed temperature, snapshot at different time periods, for [top to bottom: Aggregation of RBC agents, SARL(K), MARL(K), MARL(K,I)]}
\end{figure}

The improvement in learned transition model translates directly into greater rewards for the RL agents over time. This is visualized in Fig. 3a where RL based strategies reduce energy consumption by up to 40\%. The savings are highest for the single reinforcement learner with domain knowledge (SARL(K)); however, as mentioned above, these savings come at the cost of reduced occupant comfort (defined as number of hot water draws below 45\textdegree C) (Fig. 3b). SARL(k) is the only configuration where this boundary is breached repeatedly as seen in Fig. 3b. This is a direct consequence of learning the incorrect model of the storage vessel and a comparable multi-agent configuration does not suffer from this problem.

\begin{figure}
\includegraphics[width=1.5in]{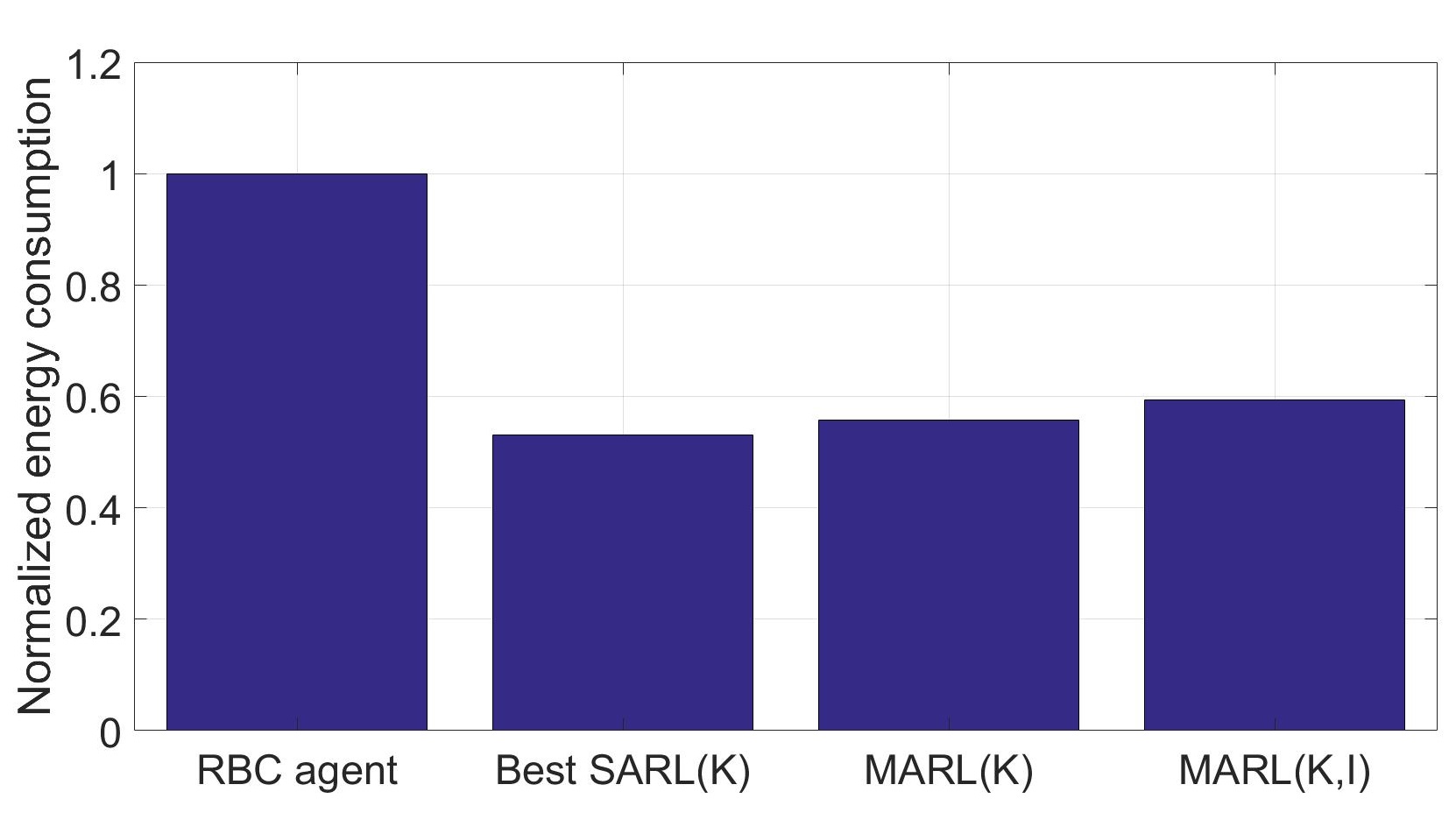}
\includegraphics[height=.9in]{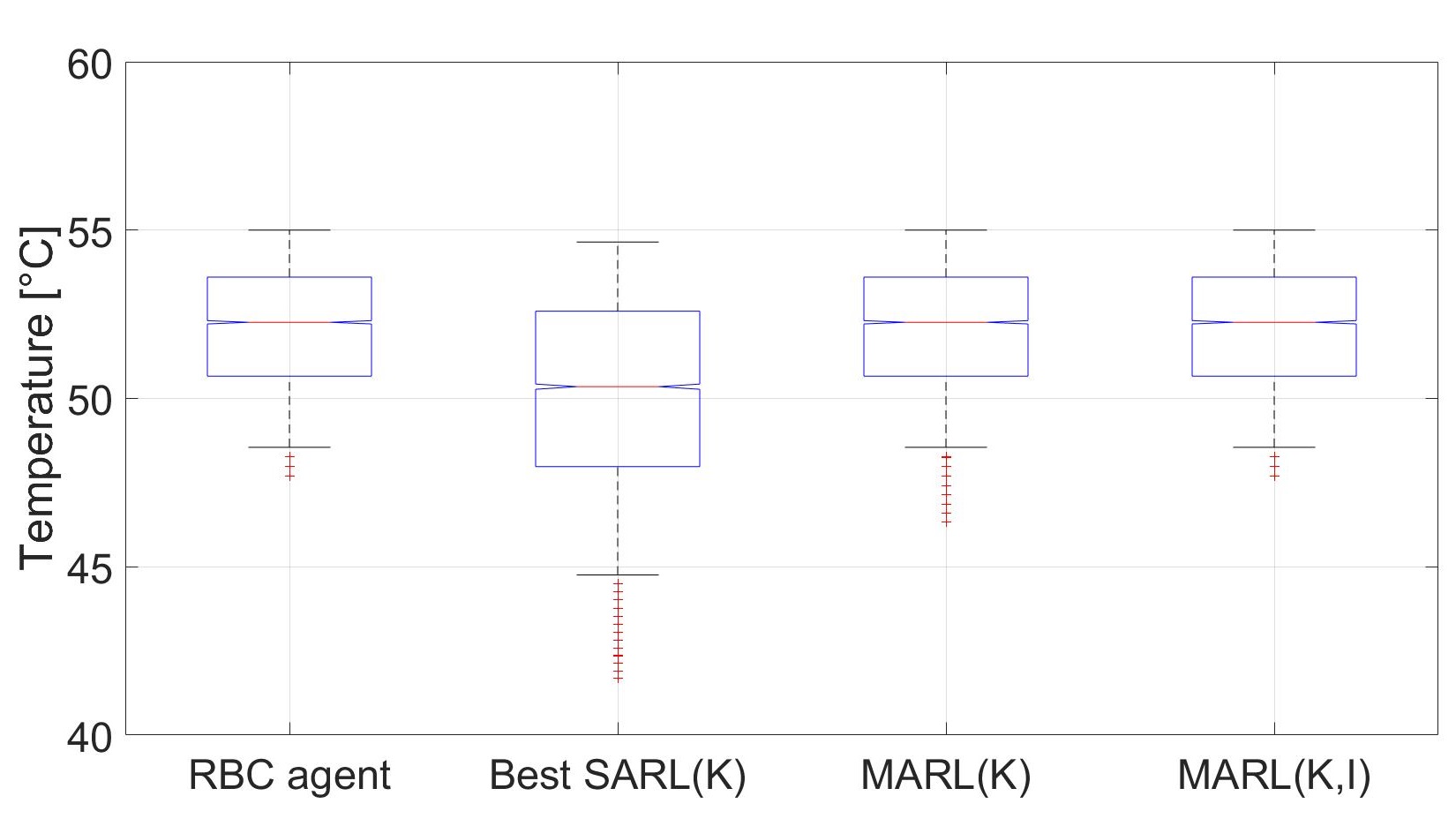}
\caption{(a) Energy consumption; (b) Water consumption temperature}
\end{figure}

\begin{acks}
This work has been carried out with support from InnoEnergy and VLAIO.
\end{acks}


\bibliographystyle{ACM-Reference-Format}  
\bibliography{sample-bibliography}  

\end{document}